\documentclass{llncs}

\usepackage{oldlfont,amssymb,epsf,stmaryrd}
\newbox\tempa
\newbox\tempb
\newdimen\tempc
\def\mud#1{\hfil $\displaystyle{\mathstrut #1}$\hfil}
\def\rig#1{\hfil $\displaystyle{#1}$}
\def\irulehelp#1#2#3{\setbox\tempa=\hbox{$\displaystyle{\mathstrut #2}$}%
                        \setbox\tempb=\vbox{\halign{##\cr
        \mud{#1}\cr
        \noalign{\vskip\the\lineskip}
        \noalign{\hrule height 0pt}
        \rig{\vbox to 0pt{\vss\hbox to 0pt{${\; #3}$\hss}\vss}}\cr
        \noalign{\hrule}
        \noalign{\vskip\the\lineskip}
        \mud{\copy\tempa}\cr}}
                      \tempc=\wd\tempb
                      \advance\tempc by \wd\tempa
                      \divide\tempc by 2 }
\def\irule#1#2#3{{\irulehelp{#1}{#2}{#3}
                     \hbox to \wd\tempa{\hss \box\tempb \hss}}}

\def\ex{\exists}
\def\fa{\forall}
\def\ra{\rightarrow}
\def\SN{{\cal SN}}
\def\lra{\longrightarrow}
\def\nulll{\mbox{\it Null\/}}
\def\pred{\mbox{\it Pred\/}}
\def\fst{\mbox{\it fst\/}}
\def\snd{\mbox{\it snd\/}}

\begin{document}
\title{Specifying programs with propositions and with congruences}
\author{Gilles Dowek}
\date{}
\institute{\'Ecole polytechnique and INRIA\\
LIX, \'Ecole polytechnique,
91128 Palaiseau Cedex, France. \\
{\tt gilles.dowek@polytechnique.edu, http://www.lix.polytechnique.fr/\~{}dowek}}
\maketitle

\section{Introduction}

Deduction modulo is an extension of first-order predicate logic where
axioms are replaced by a congruence, modulo which deduction is
performed.  Like first-order predicate logic, Deduction modulo is a
framework where many theories can be expressed, in particular {\em
predicative arithmetic} \cite{DWHA} and {\em impredicative
arithmetic}, usually called {\em first-order} and {\em second-order
arithmetic} (terminology that we try to avoid as both theories are
expressed in a first-order setting).

Krivine and Parigot's {\em Second-order functional arithmetic}
(FA$_2$) \cite{Leivant,KrivineParigot,Krivine} is a formulation of 
impredicative
arithmetic tailored to be used to specify and prove programs.  In this
note, we give a presentation of FA$_2$ in Deduction modulo. Expressing
FA$_2$ in Deduction modulo will shed light on an original aspect of
the FA$_2$ approach: the fact that programs are specified, not with
propositions, but with congruences.

\section{Deduction modulo}

\subsection{Identifying propositions}

In Deduction modulo, a theory is a congruence $\equiv$ on terms and 
propositions, often defined with rewrite rules and equational axioms. 
If ${\cal R}$ is a set of rewrite rules and equational axioms, the
congruence $\equiv_{\cal R}$ 
{\em defined by the rewrite system ${\cal R}$} 
is inductively defined as follows
\begin{itemize}
\item if $l \lra r$ is a rewrite rule or $\fa x_1 ...\fa x_n~(l = r)$
(where all the free variables of $l$ and $r$ are among $x_1, ..., x_n$)
is an 
equational axiom of ${\cal R}$ and $\sigma$ is a substitution, then 
$\sigma l \equiv_{\cal R} \sigma r$, 
\item the relation $\equiv_{\cal R}$ is reflexive, symmetric and
  transitive,
\item the relation $\equiv$ is compatible with the structure of 
terms and propositions, i.e. if $f$ is a function symbol
and $t_1 \equiv_{\cal R} u_1$, ..., 
$t_n \equiv_{\cal R} u_n$, then $f(t_1, ..., t_n) \equiv_{\cal R}
f(u_1, ..., u_n)$, and similarly for predicate symbols, connectors
and quantifiers.
\end{itemize}

The deduction rules are expressed in such 
a way that deduction is performed modulo this congruence. The rules 
of intuitionistic natural deduction modulo are given 
Fig.~\ref{NatMod}. 

\begin{figure}
\noindent\framebox{\parbox{\textwidth
}{
$$
\hspace*{-4cm}
\begin{array}{c}
\irule{}
      {\Gamma \vdash B}
      {\mbox{axiom if $A \in \Gamma$ and $A \equiv B$}}
\vspace{1.5mm}\\
\irule{}
      {\Gamma \vdash A}
      {\mbox{$\top$-intro if $A \equiv \top$}}
\vspace{1.5mm}\\
\irule{\Gamma \vdash B}
      {\Gamma \vdash A}
      {\mbox{$\bot$-elim if $B \equiv \bot$}}
\vspace{1.5mm}\\
\irule{\Gamma \vdash A~~~\Gamma \vdash B}
      {\Gamma \vdash C}
      {\mbox{$\wedge$-intro if $C \equiv (A \wedge B)$}}
\vspace{1.5mm}\\
\irule{\Gamma \vdash C}
      {\Gamma \vdash A}
      {\mbox{$\wedge$-elim if $C \equiv (A \wedge B)$}}
\vspace{1.5mm}\\
\irule{\Gamma \vdash C}
      {\Gamma \vdash B}
      {\mbox{$\wedge$-elim if $C \equiv (A \wedge B)$}}
\vspace{1.5mm}\\
\irule{\Gamma \vdash A}
      {\Gamma \vdash C}
      {\mbox{$\vee$-intro if $C \equiv (A \vee B)$}}
\vspace{1.5mm}\\
 \irule{\Gamma \vdash B}
      {\Gamma \vdash C}
      {\mbox{$\vee$-intro if $C \equiv (A \vee B)$}}
\vspace{1.5mm}\\
\irule{\Gamma \vdash D~~~\Gamma, A \vdash C~~~\Gamma, B \vdash C}
      {\Gamma \vdash C}
      {\mbox{$\vee$-elim if $D \equiv (A \vee B)$}}
\vspace{1.5mm}\\
\irule{\Gamma, A \vdash B}
      {\Gamma \vdash C}
      {\mbox{$\Rightarrow$-intro if $C \equiv (A
        \Rightarrow B)$}}
\vspace{1.5mm}\\
\irule{\Gamma \vdash C~~~\Gamma \vdash A}
      {\Gamma \vdash B}
      {\mbox{$\Rightarrow$-elim if $C \equiv (A \Rightarrow B)$}}
\vspace{1.5mm}\\
\irule{\Gamma \vdash A}
      {\Gamma \vdash B}
      {\mbox{$\langle x,A \rangle$ $\fa$-intro if $B \equiv (\fa x~A)$ and
                         $x \not\in FV(\Gamma)$}}
\vspace{1.5mm}\\
\irule{\Gamma \vdash B}
      {\Gamma \vdash C}
      {\mbox{$\langle x,A,t \rangle$ $\fa$-elim if $B \equiv (\fa x~A)$ and $C
       \equiv (t/x)A$}}
\vspace{1.5mm}\\
 \irule{\Gamma \vdash C}
        {\Gamma \vdash B}
        {\mbox{$\langle x,A,t \rangle$ $\exists$-intro if $B \equiv (\exists x~A)$ and $C
               \equiv (t/x)A$}}
\vspace{1.5mm}\\
\irule{\Gamma \vdash C~~~\Gamma, A \vdash B}
        {\Gamma \vdash B}
        {\mbox{$\langle x,A \rangle$ $\exists$-elim if $C \equiv (\exists x~A)$ and
               $x \not\in FV(\Gamma B)$}}
\end{array}$$
\caption{Natural deduction modulo \label{NatMod}}}}
\end{figure}

For instance, if the congruence $\equiv$ is defined with the rules
$$\begin{array}{rcl}
0 + y &\lra& y\\
S(x) + y &\lra& S(x+y)\\
0 \times y &\lra& 0\\
S(x) \times y &\lra& x \times y + y\\
x = x &\lra& \top
\end{array}$$
we have $(2 + 2 = 4) \equiv \top$ and we
can prove, in natural deduction modulo $\equiv$, that the number
$4$ is even 
$$\irule{\irule{}
               {\vdash 2 \times 2 = 4}
               {\mbox{$\top$-intro}}
        }
        {\vdash  \exists x~2 \times x = 4}
        {\mbox{$\langle x,2 \times x = 4,2 \rangle$ $\exists$-intro}}$$

\subsection{Proof-terms}

We sometime write natural deduction proofs modulo as proof-terms. 

\begin{definition}[Proof-term]
{\em Proof-terms} are inductively defined as follows. 

\begin{tabbing}
$\pi$ \= $= \alpha$\\
      \> $~|~I$\\
      \> $~|~\delta_{\bot}(\pi)$\\
      \> $~|~\langle \pi_1,\pi_2 \rangle~|~\fst(\pi)~|~\snd(\pi)$\\
      \> $~|~i(\pi)~|~j(\pi)~|~\delta(\pi_1,\alpha \pi_2,\beta \pi_{3})$\\ 
      \> $~|~\lambda \alpha~\pi~|~(\pi_1~\pi_2)$\\
      \> $~|~\lambda x~\pi~|~(\pi~t)$\\
      \> $~|~\langle t,\pi \rangle~|~ \delta_{\ex}(\pi_1,x \alpha \pi_2)$
\end{tabbing}
\end{definition}

Each proof-term construction corresponds to an intuitionistic natural
deduction rule: terms of the form $\alpha$ express proofs built with
the axiom rule, 
the term $I$ expresses the proof built with
the introduction rule of the symbol $\top$, terms of the form
$\delta_{\bot}(\pi)$ express proofs built with the elimination rule of
the symbol $\bot$, 
terms of the form $\langle
\pi_1,\pi_2 \rangle$ and $\fst(\pi)$, $\snd(\pi)$ express proofs built with
the introduction and elimination rules of the conjunction, terms of
the form $i(\pi), j(\pi)$ and $\delta(\pi_1,\alpha \pi_2,\beta
\pi_{3})$ express proofs built with the introduction and elimination
rules of the disjunction, 
terms of the form $\lambda \alpha~\pi$ and
$(\pi_1~\pi_2)$ express proofs built with the introduction and
elimination rules of the implication, 
terms of the form $\lambda x~\pi$ and $(\pi~t)$
express proofs built with the introduction and elimination rules of
the universal quantifier and terms of the form $\langle t,\pi \rangle$
and $\delta_{\ex}(\pi_1,x \alpha \pi_2)$ express proofs built with the
introduction and elimination rules of the existential quantifier.

\begin{definition}[Reduction]
{\em Reduction} on proofs is defined by the following rules that
eliminate cuts step by step.
$$
\begin{array}{rcl}
\fst(\langle \pi_1,\pi_2 \rangle) &\triangleright& \pi_1 \\
\snd(\langle \pi_1,\pi_2 \rangle) &\triangleright& \pi_2\\
\delta(i(\pi_1),\alpha \pi_2,\beta \pi_{3}) 
&\triangleright& (\pi_1/\alpha)\pi_2 \\
\delta(j(\pi_1),\alpha \pi_2,\beta \pi_{3}) 
&\triangleright& (\pi_1/\beta)\pi_{3}\\
(\lambda \alpha~\pi_1~\pi_2) &\triangleright&(\pi_2/\alpha)\pi_1\\
(\lambda x~\pi~t) &\triangleright& (t/x)\pi\\
\delta_{\ex}(\langle t,\pi_1 \rangle,\alpha x\pi_2) &\triangleright&
(t/x,\pi_1/\alpha)\pi_2
\end{array}$$
We write $\triangleright^*$ for the reflexive-transitive closure of
the relation $\triangleright$.
\end{definition}

A proof is said to be {\em strongly normalizing} if its reduction tree
for the relation $\triangleright$ is finite. We write
$\SN$ for the set of strongly normaling proofs.

\section{FA$_2$ in Deduction modulo}

\subsection{Classes in first-order predicate logic and in Deduction modulo}

The induction principle can be expressed as the fact that a class of 
natural numbers that contains $0$ and that is closed by the successor function 
contains all the natural numbers. A class of numbers is just a collection 
of numbers definable in comprehension by a proposition of arithmetic.

To formalize the induction principle, one possibility is to leave the
notion of class implicit and formalize this principle as an axiom scheme,
with one axiom for each proposition defining a class in
comprehension. Another possibility is to introduce a sort $\kappa$ for
classes and a binary predicate symbol $\epsilon$ for the membership of
a natural number to a class.

This way, we obtain a two-sorted theory with a sort $\iota$ for natural 
numbers and a sort $\kappa$ for classes and we can formulate the induction 
axiom as the proposition 
$$\fa X~(0~\epsilon~X \Rightarrow \fa y~(N(y) \Rightarrow y~\epsilon~X \Rightarrow S(y)~\epsilon~X) \Rightarrow \fa n~(N(n) \Rightarrow n~\epsilon~X))$$
To be able define classes in comprehension, for instance to define the class
of 
the elements 
$x$ such that $\ex w~(x = 2 \times w)$, we need an axiom 
expressing the existence of this class. Thus, for each proposition $A$,
we need an axiom expressing that there exists a class $Z$ 
such that $x~\epsilon~Z$ if and only if $A$
$$\fa y_1 \ldots \fa y_p \ex Z \fa x~(x~\epsilon~Z \Leftrightarrow A)$$
where the free variables of $A$ are among $x, y_1, \ldots, y_p$. 

In this comprehension scheme, we must be precise about the
propositions $A$ that define instances of the scheme. If we want to
have equivalence with the ``usual'' formulation of predicative
arithmetic, $A$ must be a proposition in the language we had before
adding the sort $\kappa$ and the predicate symbol $\epsilon$. In
particular, $A$ must not contain quantifiers over variables of sort
$\kappa$.

Skolemizing each of these axioms, introduces an infinite number of 
Skolem symbols $f_{x, y_1, \ldots, y_p, A}$ and the universal axiom 
$$\fa y_1 \ldots \fa y_p \fa x~(x~\epsilon~f_{x, y_1, \ldots, y_p, A}(y_1, \ldots, y_p) 
\Leftrightarrow A)$$
In Deduction modulo, these axioms can be transformed into rewrite rules 
$$x~\epsilon~f_{x,y_1, \ldots, y_p, A}(y_1, \ldots, y_p) \lra A$$

This construction generalizes to $n$-ary classes: for each $n$, 
we introduce a sort $\kappa_n$ for $n$-ary classes, a $(n+1)$-ary predicate
symbol $\epsilon_n$, a $n$-ary comprehension scheme 
$$\fa y_1 \ldots \fa y_p \ex Z \fa x_1  \ldots \fa x_n~(\epsilon_n(x_1, \ldots, x_n,Z)
\Leftrightarrow A)$$
We can skolemize this scheme and introduce 
Skolem symbols $f_{x_1, \ldots, x_n, y_1, \ldots, y_p, A}$. Finally, we can
transform
the skolemized axioms into rewrite rules
$$\epsilon_n(x_1, \ldots, x_n,f_{x_1, \ldots, x_n ,y_1, \ldots, y_p, A}(y_1,
 \ldots, y_p))
 \lra A$$

Following an idea of Von Neuman, Bernays and G\"odel, but using the more
recent tool of explicit substitutions \cite{ACCL}, Kirchner has shown that 
this infinite number of Skolem symbols, could be replaced by a finite
language \cite{Florent}, but, we shall not use this here.

\subsection{Arithmetic in Deduction modulo}

Together with Werner, we have given in \cite{DWHA} a 
presentation of predicative arithmetic in Deduction modulo. Starting with
the theory of unary classes, we have transformed Leibniz definition of 
equality into a rewrite rule 
$$y = z \lra \fa X~(y~\epsilon~X \Rightarrow z~\epsilon~X)$$
Using trivial equivalences in predicate logic, the induction axiom above
can be reformulated as 
$$\fa n~(N(n) \Rightarrow \fa X~(0~\epsilon~X \Rightarrow 
\fa y~(N(y) \Rightarrow y~\epsilon~X \Rightarrow S(y)~\epsilon~X) \Rightarrow
n~\epsilon~X))$$
Then using the class $f_{x, N(x)}$ and the Peano
first and second axioms ($N(0)$ and 
$\fa x~(N(x) \Rightarrow N(S(x)))$
this implication can be transformed into an equivalence
$$\fa n~(N(n) \Leftrightarrow \fa X~(0~\epsilon~X \Rightarrow 
\fa y~(N(y) \Rightarrow y~\epsilon~X \Rightarrow S(y)~\epsilon~X) \Rightarrow
n~\epsilon~X))$$
and then into a rewrite rule 
$$N(n) \lra   \fa X~(0~\epsilon~X \Rightarrow \fa
  y~(N(y) \Rightarrow y~\epsilon~X \Rightarrow S(y)~\epsilon~X) \Rightarrow n~\epsilon~X)$$
The Peano third and fourth axiom ($\fa x \fa y~(S(x) = S(y) \Rightarrow x = y)$ and $\fa x \neg S(x) = 0$) can be transformed into rewrite rules if we 
introduce a new function symbol $\pred$ and a new predicate symbol $\nulll$
$$\begin{array}{rcl}
\pred(0) &\lra& 0\\
\pred(S(x)) &\lra& x\\
\nulll(0) &\lra& \top\\
\nulll(S(x)) &\lra& \bot
\end{array}$$

Finally, the axioms of addition and multiplication can be trivially 
transformed into rewrite rules. 

We get this way the rewrite system $HA$ of Fig.~\ref{arith}, that defines
a congruence $\equiv_{HA}$ and we
can prove that Deduction modulo $\equiv_{HA}$ is a conservative extension of
intuitionistic predicative arithmetic \cite{DWHA}. This system can be
divided into two subsystems $HA_1$ contains the five rules that
rewrite atomic propositions to propositions and $HA_2$ the six rules
that rewrite terms to terms.

\begin{figure}[!t]
\noindent\framebox{\parbox{\textwidth
}{
$$x~\epsilon~f_{x, y_1, \ldots, y_n, A}(y_1, \dots, y_n) \lra A$$
$$y = z \lra \fa X~(y~\epsilon~X \Rightarrow z~\epsilon~X)$$
$$N(n) \lra   \fa X~(0~\epsilon~X \Rightarrow \fa
  y~(N(y) \Rightarrow y~\epsilon~X \Rightarrow S(y)~\epsilon~X) 
\Rightarrow n~\epsilon~X)$$
$$\nulll(0) \lra \top$$
$$\nulll(S(x)) \lra \bot$$
$$\begin{array}{rclrcl}
\pred(0) &\lra& 0~~~~~~~~~~~ & \pred(S(x)) &\lra& x\\
0 + y &\lra& y  & S(x) + y &\lra& S(x+y)\\
0 \times y &\lra& 0 &  S(x) \times y &\lra& x \times y + y
\end{array}$$
\caption{The rewrite system $HA$\label{arith}}
}}
\end{figure}

\subsection{Impredicative arithmetic in Deduction modulo}

Extending this theory to impredicative (i.e. second-order) 
arithmetic just requires to extend the comprehension scheme and
introduce symbols $f_{x, y_1, \ldots, y_n, A}$ and the associated rewrite
rules for all propositions $A$, including those containing the symbol
$\epsilon$, variables of sort $\kappa$ and quantifiers over these variables.

\subsection{Datatypes}

This theory can be extended to other datatypes such as lists
or trees: starting from the theory of unary classes, we introduce 
function symbols for the constructors of the datatype, a unary
predicate symbol for the membership to the datatype, function symbols for the 
right-inverses of the constructors and predicate symbols for the images of the 
constructors. Then, the associated rewrite rules permit to prove the closure
of the datatype by the constructors (the equivalents of the Peano 
first and second axioms), the induction principle (the equivalent of the 
Peano fifth axiom),
the injectivity of constructors (the equivalent of the Peano third axiom) and 
the non confusion of the constructors (the equivalent of the Peano fourth 
axiom). 

\section{Normalization}

Poofs in Deduction modulo do not normalize for
all theories.
For instance, 
taking the rule $P \lra (Q \Rightarrow P)$, we obtain a theory where
all proofs normalize, but
taking the rule 
$P \lra (P \Rightarrow Q)$, 
we obtain a theory where proofs do not always normalize.

Together with Werner, we have proved in \cite{DW} that all proofs
normalize modulo some congruence if this congruence has a ${\cal
C}$-model, i.e. a model where propositions are valued, not in the
algebra $\{0,1\}$, but in the algebra ${\cal C}$ of reducibility
candidates. We now define this algebra.

\subsection{Reducibility candidates}

\begin{definition}[Reducibility candidates]\label{defCR}\cite{Girard}
A proof is said to be {\em neutral} 
if it is an axiom or an elimination (i.e. of the form 
$\alpha$, $\delta_{\bot}(\pi)$, 
$\fst(\pi)$, $\snd(\pi)$, $\delta(\pi_1,\alpha
\pi_2,\beta \pi_{3})$, 
$(\pi~\pi')$, $(\pi~t)$, or
$\delta_{\ex}(\pi, x \alpha \pi')$), but not an introduction. A set
$R$ of proofs is a {\em reducibility candidate} if 
\begin{itemize}
\item whenever $\pi \in R$, then $\pi$ is strongly normaling,
\item whenever $\pi \in R$ and $\pi \triangleright^* \pi'$,
then $\pi' \in R$, 
\item whenever $\pi$ is neutral and 
if for every $\pi'$ such that $\pi \triangleright \pi'$, $\pi' \in R$ then 
$\pi \in R$. 
\end{itemize}
\end{definition}

We write ${\cal C}$ for the set of reducibility candidates and
we define the following operations on this set: we let
$\tilde{\top} = \tilde{\bot} = \SN$, if $A$ and $B$ are reducibility 
candidates, we let 
$$A~\tilde{\wedge}~B =\{\pi\in\SN~|~\pi \triangleright^* 
\langle \pi_1,\pi_2\rangle\Rightarrow(\pi_1\in A\wedge\pi_2\in B)\}$$
$$A~\tilde{\vee}~B =\{\pi\in\SN~|~\pi\triangleright^*
i(\pi_1)\Rightarrow\pi_1\in A \wedge\pi\triangleright^*
j(\pi_2)\Rightarrow\pi_2\in B \}$$
$$A~\tilde{\Rightarrow}~B = 
\{\pi\in\SN~|~\pi\triangleright^*\lambda\alpha~\pi'\Rightarrow
\fa\sigma\in A~(\sigma/\alpha)\pi' \in B\}$$
if $A$ is a set of reducibility candidates, we let 
$$\tilde{\fa}~A =\{\pi\in\SN~|~\pi\triangleright^*\lambda
x~\pi'\Rightarrow \fa X\in A \fa t \in {\cal T}~(t/x)\pi'\in A \}$$
where ${\cal T}$ is the set of terms of the language and
$$\tilde{\exists}~A = \{\pi\in\SN~|~\pi\triangleright^*\langle
t,\pi'\rangle\Rightarrow \exists X \in A ~\pi'\in X \}$$
It is routine to check that these sets are reducibility candidates
\cite{DW}.

\subsection{${\cal C}$-models}

\begin{definition}[${\cal C}$-model]
A {\em ${\cal C}$-model} ${\cal M}$ for a many-sorted language ${\cal L}$ is
given by
\begin{itemize}
\item for every sort $s$ a set $M_s$, 
\item for every function symbol $f$ of rank $\langle
s_1,\dots,s_n,s_{n+1} \rangle$ a mapping $\hat{f}$  
from $M_{s_1} \times \dots \times M_{s_n}$ to $M_{s_{n+1}}$, 
\item for every predicate symbol $P$ of rank  $\langle s_1,\dots,s_n \rangle$ a
mapping $\hat{P}$ 
from $M_{s_1} \times \dots \times M_{s_n}$ to ${\cal C}$.
\end{itemize}
\end{definition}

The denotation of a term or a proposition in a ${\cal C}$-model is defined
as follows. 

\begin{itemize}
\item $\llbracket x\rrbracket_{\phi} = \phi(x)$, 
$\llbracket f(t_1, \ldots, t_n)\rrbracket_{\phi} = \hat{f}(\llbracket t_1\rrbracket_{\phi},
\ldots, \llbracket t_n\rrbracket_{\phi})$,
\item $\llbracket P(t_1, \ldots, t_n)\rrbracket_{\phi} =
\hat{P}(\llbracket t_1\rrbracket_{\phi}, \ldots, \llbracket
t_n\rrbracket_{\phi})$, 
\item $\llbracket \top \rrbracket_{\phi} = \tilde{\top}$, 
$\llbracket \bot \rrbracket_{\phi} = \tilde{\bot}$, 
\item $\llbracket A \wedge B \rrbracket_{\phi} = 
\llbracket A \rrbracket_{\phi}~\tilde{\wedge}~
\llbracket B \rrbracket_{\phi}$, 
$\llbracket A \vee B \rrbracket_{\phi} = 
\llbracket A \rrbracket_{\phi}~\tilde{\vee}~
\llbracket B \rrbracket_{\phi}$, 
$\llbracket A \Rightarrow B \rrbracket_{\phi} = 
\llbracket A \rrbracket_{\phi}~\tilde{\Rightarrow}~
\llbracket B \rrbracket_{\phi}$, 
\item 
$\llbracket \fa x~A \rrbracket_{\phi} = 
\tilde{\fa}\{\llbracket A \rrbracket_{\phi + a/x}~|~a \in M_s\}$, 
$\llbracket \ex x~A \rrbracket_{\phi} = 
\tilde{\ex}\{\llbracket A \rrbracket_{\phi + a/x}~|~a \in M_s\}$. 
\end{itemize}

A ${\cal C}$-model is said to be a {\em ${\cal C}$-model of a 
congruence $\equiv$}
if when $A \equiv B$ then for 
every assignment $\phi$, $\llbracket A \rrbracket_{\phi} = \llbracket
B \rrbracket_{\phi}$.  

\begin{proposition}
If a congruence has a ${\cal C}$-model, then all proofs modulo this congruence
strongly normalize.
\end{proposition}

\proof{See \cite{DW}.}

\subsection{Normalization for arithmetic}

We can build a ${\cal C}$-model for both predicative and impredicative
arithmetic by taking $M_{\iota} = {\mathbb N}$ and $M_{\kappa} =
{\mathbb N} \ra {\cal C}$ (the set of functions from ${\mathbb N}$ to
${\cal C}$).  The symbols $0$, $S$, $+$, $\times$, and $\pred$ are
interpreted in the standard way. The symbol $\epsilon$ is interpreted
by the function mapping $n$ and $f$ to $f(n)$.  We define
$\hat{\nulll}$ as the function mapping $0$ to $\tilde{\top}$ and the
other numbers to $\tilde{\bot}$.  The equality predicate, the predicate $N$
and the Skolem symbols are interpreted in such a way that the
corresponding rewrite rules are valid \cite{DWHA}.

However, we shall use another ${\cal C}$-model that will be more useful 
later, where $M_{\iota}$ is a 
singleton. With usual models, where propositions are valued in
$\{0,1\}$, 
the set $M_{\iota}$ cannot be a singleton
because the rules 
$\nulll(0) \lra \top$ and $\nulll(S(x)) \lra \bot$,
that replace the Peano fourth
axiom, require $\hat{\nulll}(\hat{0}) = 1$ and 
$\hat{\nulll}(\hat{S}(\hat{0})) = 0$ and thus 
$\hat{0}$ and $\hat{S}(\hat{0})$ to be distinct.
For models where propositions are valued in ${\cal C}$, we have the conditions 
$\hat{\nulll}(\hat{0}) = \tilde{\top}$ and 
$\hat{\nulll}(\hat{S}(\hat{0})) = \tilde{\bot}$, but 
$\tilde{\top} = \tilde{\bot} = \SN$, thus nothing prevents 
$M_{\iota}$ from being a singleton. 

\begin{proposition}\label{normHA}
All proofs normalize modulo $\equiv_{HA}$.
\end{proposition}

\proof{We construct a ${\cal C}$-model as follows.

We let $M_{\iota}$ be a singleton $\{0\}$ and $M_{\kappa} = {\cal C}$.

The symbols $0$, $S$, $+$, $\times$, and $\pred$ are
interpreted in the trivial way. 

The symbol $\epsilon$ is interpreted
by the function mapping $0$ and $C$ to $C$.

We define $\hat{=}$ as the function mapping $0$ and $0$ to 
$\llbracket \fa X~(y~\epsilon~X \Rightarrow
z~\epsilon~X)\rrbracket_{0/y,0/z}$.

To define $\hat{N}$ we first define the function $\Phi$ that maps any
reducibility candidate $C$ to the interpretation of the proposition
$\fa X~(0~\epsilon~X \Rightarrow \fa y~(N(y) \Rightarrow y~\epsilon~X \Rightarrow S(y)~\epsilon~X) \Rightarrow n~\epsilon~X)$
in the model of domains $M_{\iota}$ and $M_{\kappa}$ and
where $0$ is interpreted by $\hat{0}$, $S$ by $\hat{S}$,
$\epsilon$ by $\hat{\epsilon}$, 
but $N$ by $C$. 
The set of reducibility candidates ordered by inclusion is complete and 
the function $\Phi$ is monotonous, thus it has a fixed point $D$.
We let $\hat{N} = D$. 

We define $\hat{\nulll}$ as the function mapping $0$ to $\tilde{\top}
= \tilde{\bot}$.

This way we can interpret all the propositions $A$ that do not
contain Skolem symbols. 
Finally, we interpret the symbols 
$f_{x, y_1, \ldots, y_n, A}$ as the functions mapping 
$0, \ldots, 0$ to 
$\llbracket A \rrbracket_{0/x,0/y_1, \ldots, 0/y_n}$.} 

\section{Non confusion and uniformity}

As usual, we want to prove that normal closed proofs are {\em uniform}, 
i.e. that they end with an introduction rule. This requires an extra 
property of the congruence: that it is {\em non confusing}. 

A congruence $\equiv$ is said to be {\em non confusing} when if two non atomic 
propositions $A$ and $B$ are congruent, then either 
\begin{itemize}
\item $A = B = \top$, 
\item $A = B = \bot$, 
\item $A = A_1 \wedge A_2$, $B = B_1 \wedge B_2$, $A_1 \equiv B_1$ and 
$A_2 \equiv B_2$, 
\item $A = A_1 \vee A_2$, $B = B_1 \vee B_2$, $A_1 \equiv B_1$ and 
$A_2 \equiv B_2$, 
\item $A = A_1 \Rightarrow A_2$, $B = B_1 \Rightarrow B_2$, 
$A_1 \equiv B_1$ and $A_2 \equiv B_2$, 
\item $A = \fa x~A_1$, $B = \fa x~B_1$ and $A_1 \equiv B_1$, 
\item $A = \ex x~A_1$, $B = \ex x~B_1$ and $A_1 \equiv B_1$. 
\end{itemize}

The following Proposition allows to prove, for instance, that
$\equiv_{HA}$ is non  confusing.

\begin{proposition}
If a congruence is defined by a confluent rewrite system that rewrite terms
to terms and atomic propositions to propositions, then it is non confusing.
\end{proposition}

\proof{If $A$ and $B$ are non atomic propositions such that $A \equiv B$,
then by confluence, there exits a proposition $C$ such that $A$ reduces to 
$C$ and $B$ reduces to $C$. Thus $C$ is non atomic, the main connector or
quantifier of $A$ and $B$ are the same as that of $C$ and the components of 
$A$ and $B$ rewrite to those of $C$.}

\begin{proposition}
If a congruence $\equiv$ is non confusing, then normal closed proofs end 
with an introduction rule.
\end{proposition}

\proof{By induction over proof structure.}

\section{Specifying programs}

Deduction modulo $\equiv_{HA}$ can be used in at least two different ways
to specify programs. 

\subsection{Proofs as programs}

If $n$ is a natural number, we write 
$\underline{n}$ for the term $S^n(0)$ and we write $\rho_n$ for the
{\em Parigot numeral} $n$ \cite{Parigot}, i.e. the proof of $N(\underline{n})$
defined as follows
\begin{itemize}
\item $\rho_0 = \lambda X \lambda \alpha \lambda \beta~\alpha$
\item $\rho_{n+1} = 
\lambda X \lambda \alpha \lambda
\beta~(\beta~\underline{n}~\rho_n~(\rho_n~X~\alpha~\beta))$. 
\end{itemize}

\begin{proposition}
Let $t$ be a term and $\pi$ be a normal closed proof of $N(t)$ modulo
$\equiv_{HA}$, then there exists a natural number $n$ such that $\pi = \rho_n$
and $t \equiv_{HA} \underline{n}$. 
\end{proposition}

\proof{By induction on the structure of $\pi$.}

\medskip

If $\pi$ is a closed proof of the proposition
$$\fa x_1~(N(x_1) \Rightarrow \fa x_2~(N(x_2) 
\Rightarrow \ldots \Rightarrow \fa x_n~(N(x_n) \Rightarrow \ex
y~N(y))\ldots))$$
then for each $p_1, \ldots, p_n$ the normal form of the proof 
$(\pi~\underline{p}_1~\rho_{p_1}~\underline{p}_2~\rho_{p_2}~\ldots~
\underline{p}_n~\rho_{p_n})$ is a pair $\langle t, \sigma \rangle$
where $t$ is a term and $\sigma$ a normal closed proof of $N(t)$.
Thus, there exists a natural number $q$ such that $\sigma = \rho_q$ and 
$t \equiv_{HA} \underline{q}$. 

The proof $\pi$ is a program mapping $p_1, \ldots, p_n$ to $q$. 

\subsection{Specifying programs with propositions}

The first method to specify programs is to use the fact that for every 
computable partial
function $f$ from ${\mathbb N}^n$ to ${\mathbb N}$, there exists a 
proposition $A$ such that 
the proposition $(\underline{p}_1/x_1, \ldots, \underline{p}_n/x_n,\underline{q}/y)A$ is
provable in arithmetic
if and only if 
$q = f(p_1, \ldots, p_n)$.

\begin{example}
The characteristic function of the set of even numbers is represented 
by the proposition
$$((\ex z~x = 2 \times z) \Rightarrow y = 1) \wedge 
((\ex z~x = 2 \times z + 1) \Rightarrow y = 0)$$
\end{example}

If $\pi$ is a proof of the proposition 
$$\fa x_1~(N(x_1) \Rightarrow \fa x_2~(N(x_2) \Rightarrow \ldots
\Rightarrow \fa
x_n~(N(x_n) \Rightarrow \ex y~(N(y) \wedge A))\ldots))$$
then the normal form of the proof 
$(\pi~\underline{p}_1~\rho_{p_1}~\underline{p}_2~\rho_{p_2}~\ldots~
\underline{p}_n~\rho_{p_n})$ is a pair $\langle t, \langle \sigma, \tau
\rangle \rangle$
where $t$ is a normal term and $\sigma$ a proof of $N(t)$ and $\tau$ a 
proof of 
$(\underline{p}_1/x_1, \ldots, \underline{p}_n/x_n,t/y)A$.

Thus, there exists a natural number $q$ such that $t \equiv_{HA}
\underline{q}$ and $\sigma = \rho_q$. The proposition 
$(\underline{p}_1/x_1, \ldots, \underline{p}_n/x_n,\underline{q}/y)A$ is 
provable hence $q = f(p_1, \ldots, p_n)$. 

The proof 
$$\pi' = \lambda x_1 \lambda \alpha_1 \ldots \lambda x_n \lambda
\alpha_n~
\langle fst (\pi~x_1~\alpha_1~\ldots~x_n~\alpha_n),
\fst (\snd (\pi~x_1~\alpha_1~\ldots~x_n~\alpha_n)) \rangle$$ 
is a proof of 
$$\fa x_1~(N(x_1) \Rightarrow \fa x_2~(N(x_2) \Rightarrow \ldots
\Rightarrow \fa x_n~(N(x_n) \Rightarrow \ex y~N(y))\ldots))$$
and it is a program mapping 
$p_1, \ldots, p_n$ to $f(p_1, \ldots, p_n)$.

\subsection{Specifying programs with congruences}

An original aspect of the FA$_2$ approach is that the programs are 
not specified as above, using a proposition representing the function in 
arithmetic, but using a set of equations {\em \`a la} Herbrand-G\"odel. 
Recall that for each 
computable function $f$, there exists a function symbol $F_0$, 
auxiliary function symbols $F_1, \ldots, F_k$ and 
a set of equational axioms ${\cal E}$ such that
the proposition $\underline{q} = 
F_0(\underline{p}_1, \ldots, \underline{p}_n)$ is provable from the axioms 
of ${\cal E}$ and the axioms of equality
if and only if
$q = f(p_1, \ldots, p_n)$.

\begin{example}
The characteristic function of the set of even numbers is represented 
by the equations
$F(0) = S(0)$, $F(S(0)) = 0$, $\fa x~(F(S(S(x))) = F(x))$.
\end{example}

In the FA$_2$ approach the equalities of ${\cal E}$ do not have any 
computational content: if $t = u$ is such an equation, and $A$ an arbitrary 
proposition, any 
proof of $(t/x)A$ is a proof of $(u/x)A$ and vice versa. This can be 
formalized in Deduction modulo, as the fact that the equations of ${\cal
E}$ are part of the definition of the congruence. 

Let $\equiv_{HA \cup {\cal E}}$ be the congruence defined by the rules
of $HA$ and the equations of ${\cal E}$. 

\begin{proposition}
Deduction modulo $\equiv_{HA \cup {\cal E}}$ has the cut elimination property.
\end{proposition}

\proof{The rules $HA$ and the axioms of ${\cal E}$ are
valid in the ${\cal C}$-model build in the proof of Proposition
\ref{normHA}.}

\medskip

Non confusion, in contrast, requires an extra hypothesis on the
equations of ${\cal E}$. Indeed, if we had in ${\cal E}$ the equation
$0 = S(0)$, or even $S(0) = S(S(F(0)))$ we we would have $\nulll(0)
\equiv_{HA \cup {\cal E}} \nulll(S(0))$, or $\nulll(0) \equiv_{HA \cup
{\cal E}} \nulll(S(F(0)))$, and thus $\top \equiv_{HA \cup {\cal E}}
\bot$.

Thus, we restrict to the case where the rules and equations of $HA_2
\cup {\cal E}$ do not allow to prove a proposition of the form $0 =
S(t)$, i.e. such that the congruence $\equiv_{HA_2 \cup {\cal E}}$
verifies the property $0 \not\equiv_{HA_2 \cup {\cal E}} S(t)$. This
includes all the cases where ${\cal E}$ can be interpreted in the
standard model of arithmetic, but also others such those where 
${\cal E}$ contains the 
equation $F(0) = S(F(0))$ that correspond to  partial computable functions
not defined in $0$.

We consider the rewriting relation defined by the rules of $HA_1$
modulo $\equiv_{HA_2 \cup {\cal E}}$, it is defined by 
$A \lra_{HA_1 / HA_2 \cup {\cal E}} B$ if there exists 
$A'$ and $B'$ such that $A \equiv_{HA_2 \cup {\cal E}} A'$, 
$A' \lra_{HA_1} B'$, and 
$B' \equiv_{HA_2 \cup {\cal E}} B$, where $A' \lra_{HA_1} B'$ means that 
$A'$ rewrites to $B'$ with one of the rules of $HA_1$. 

The next proposition shows that the use of the congruence in
$\lra_{HA_1 / HA_2 \cup {\cal E}}$ can be localized in the sense of 
\cite{PetersonStickel,JouannaudKirchner}. 

We use the following notation, if $A$ is a tree and $o$ an
occurrence in $A$, we write $A_{|o}$ for the subtree of $A$ taken at
occurrence $o$ and $A[B]_o$ for the tree $A$ where the tree $B$ has
been grafted at the occurrence $o$. 

\begin{proposition}
\label{localisation}
If $A \lra_{HA_1 / HA_2 \cup {\cal E}} B$ then there exists an occurrence $o$ of an atomic 
proposition in $A$, a proposition rewrite rule $l \lra r$ of $HA_1$ and a 
substitution $\sigma$ such that 
$A_{|o} \equiv_{HA_2 \cup {\cal E}} \sigma l$ and $B \equiv_{HA_2 \cup {\cal E}} A[\sigma r]_o$. 
\end{proposition}

\proof{By induction on the structure of $A$.}

\begin{proposition}
\label{confluence}
The relation $\lra_{HA_1 / HA_2 \cup {\cal E}}$ is confluent.
\end{proposition}

\proof{We prove more generally that this relation is 
strongly confluent, i.e. if 
$A \lra_{HA_1 / HA_2 \cup {\cal E}} B_1$ and $A \lra_{HA_1 / HA_2 \cup {\cal E}} B_2$, then 
there exists a $C$ such that
($B_1 = C$ or $B_1 \lra_{HA_1 / HA_2 \cup {\cal E}} C$) and 
($B_2 = C$ or $B_2 \lra_{HA_1 / HA_2 \cup {\cal E}} C$). 

If $A \lra_{HA_1 / HA_2 \cup {\cal E}} B_1$ and 
$A \lra_{HA_1 / HA_2 \cup {\cal E}} B_2$, then by Proposition 
\ref{localisation}, there exists an
occurrence $o_1$, a rule $l_1 \lra r_1$ of $HA_1$, and a substitution 
$\sigma_1$
such that $A_{|o_1} \equiv_{HA_2 \cup {\cal E}} \sigma_1 l_1$ and $B_1
\equiv_{HA_2 \cup {\cal E}} A[\sigma_1 r_1]_{o_1}$ and an occurrence
$o_2$, a rule $l_2 \lra r_2$ of $HA_1$, and a substitution $\sigma_2$ such that
$A_{|o_2} \equiv_{HA_2 \cup {\cal E}} \sigma_2 l_2$, and $B_2
\equiv_{HA_2 \cup {\cal E}} A[\sigma_2 r_2]_{o_2}$.

If the occurrences $o_1$ and $o_2$ are distinct, they are disjoint and 
we can close the diagram in one step. 

If they are equal, then, as $0 \not\equiv_{HA_2 \cup {\cal E}} S(t)$, the rule is the same
and $B_1 \equiv_{HA_2 \cup {\cal E}} B_2$.}

\begin{proposition}
The congruence $\equiv_{HA \cup {\cal E}}$ is non confusing.
\end{proposition}

\proof{The reflexive-symmetric-transitive closure of the relation 
$\lra_{HA_1 / HA_2 \cup {\cal E}}$ is the relation $\equiv_{HA \cup {\cal E}}$.
Thus, by Proposition \ref{confluence}, if $A \equiv_{HA \cup {\cal E}} B$, 
there exists a proposition $C$ such that $A \lra_{HA_1 / HA_2 \cup {\cal E}}^* C$ and
$B \lra_{HA_1 / HA_2 \cup {\cal E}}^* C$. 
Thus $C$ is non atomic, the main connector or
quantifier of $A$ and $B$ are the same as that of $C$ and the components of 
$A$ and $B$ rewrite to those of $C$.} 

\begin{proposition}
Let $t$ be a term and $\pi$ be a normal closed proof of $N(t)$ modulo
$\equiv_{HA \cup {\cal E}}$, then there exists a natural number $n$
such that $\pi = \rho_n$ and $t \equiv_{HA \cup {\cal E}} \underline{n}$. 
\end{proposition}

\proof{By induction on the structure of $\pi$.}

\medskip

Finally, if $\pi$ is a proof of the proposition 
$$\fa x_1~(N(x_1) \Rightarrow \fa x_2~(N(x_2) \Rightarrow \ldots \Rightarrow
\fa
x_n~(N(x_n) \Rightarrow \ex y~(N(y) \wedge y = F_0(x_1, \ldots,
x_p)))\ldots))$$
then 
the normal form of the proof 
$(\pi~\underline{p}_1~\rho_{p_1}~\underline{p}_2~\rho_{p_2}~\ldots~
\underline{p}_n~\rho_{p_n})$ is a pair $\langle t, \langle \sigma, \tau
\rangle \rangle$
where $t$ is a normal term and $\sigma$ a proof of $N(t)$ and $\tau$ a 
proof of 
$t = F(\underline{p}_1, \ldots, \underline{p}_n)$. 

Thus there exists a natural number $q$, such that $\sigma = \rho_q$
and $t \equiv_{HA \cup {\cal E}} \underline{q}$ 
the proposition 
$\underline{q} = F(\underline{p}_1, \ldots, \underline{p}_n)$
is provable modulo $\equiv_{HA \cup {\cal E}}$. 

Let $r = f(p_1, \ldots, p_n)$. The proposition 
$\underline{r} = F(\underline{p}_1, \ldots, \underline{p}_n)$
is provable from the axioms of ${\cal E}$ and the axioms of 
equality, thus it is provable modulo $\equiv_{HA \cup {\cal E}}$. 

Thus, the proposition $\underline{r} = \underline{q}$ is provable modulo 
$\equiv_{HA \cup {\cal E}}$. As $\equiv_{HA \cup {\cal E}}$ is
normalizing and non confusing it is consistent, thus
$q = r$ and $q = f(p_1, \ldots, p_n)$. 

The proof 
$$\pi' = \lambda x_1 \lambda \alpha_1 \ldots \lambda x_n \lambda
\alpha_n~
\langle fst (\pi~x_1~\alpha_1~\ldots~x_n~\alpha_n),
\fst (\snd (\pi~x_1~\alpha_1~\ldots~x_n~\alpha_n)) \rangle$$ 
is a proof of 
$$\fa x_1~(N(x_1) \Rightarrow \fa x_2~(N(x_2) \Rightarrow \ldots \Rightarrow \fa
x_n~(N(x_n) \Rightarrow \ex y~N(y))\ldots))$$
and it is a program mapping 
$p_1, \ldots, p_n$ to $f(p_1, \ldots, p_n)$.

\section{Specifications} 

We have seen that the characteristic function of even numbers could be 
specified either with the proposition
$$((\ex z~x = 2 \times z) \Rightarrow y = 1) \wedge 
((\ex z~x = 2 \times z + 1) \Rightarrow y = 0)$$
or with the equational axioms
$$F(0) = S(0)~~~~~F(S(0)) = 0~~~~~\fa x~(F(S(S(x))) = F(x))$$

We may discuss if this equational specification should be called a
specification or a program as, if we orient the equations, we get the 
rewrite rules 
$$F(0) \lra S(0)~~~~~F(S(0)) \lra 0~~~~~F(S(S(x))) \lra F(x)$$
that define an algorithm computing the value of this function.
However, the situation is not always as simple. For instance, if we add
the equation $F(0) = F(0)$, we obtain a rewrite system that is not 
locally deterministic, as the set term $F(0)$ reduces both to 
$S(0)$ and to itself and that is only weakly terminating. Thus although the 
specification remains executable, its execution mechanisms is better 
described as a proof search process in an equational theory 
rather than as a mere reduction.
In this sense, the propositional specification also is executable,
as for every $p_1, ..., p_n$ we can use a proof search system to find a proof 
in arithmetic of $\ex y~(\underline{p}_1/x_1, ...,
\underline{p}_n/x_n)A$ and extract a witness
from this proof. The specification is just easier to execute in the
second case because proof search is easier in an equational theory 
than in arithmetic.
Thus, the difference between these two ways of specifying
programs is not in the fact that one specification is executable and
not the other. 

The difference seems to be that the two approaches are based on a
different notion of definition. In set theory, for instance, the
notion of intersection of two sets can either be {\em explicitly
defined} by stating that $A \cap B$ is an abbreviation for $\{x \in
A~|~x \in B\}$ or {\em implicitly defined} by introducing a new
primitive symbol $\cap$ and the axiom $x \in A \cap B \Leftrightarrow
(x \in A \wedge x \in B)$. In the same way, we can say that the
characteristic function of the set of even numbers can either be
explicitly defined by the proposition $((\ex z~x = 2 \times z)
\Rightarrow y = 1) \wedge ((\ex z~x = 2 \times z + 1) \Rightarrow y =
0)$ or implicitly defined with the introduction of a new primitive
symbol $F$ and the axioms $F(0) = S(0)$, $F(S(0)) = 0$, $\fa
x~(F(S(S(x))) = F(x))$.

Using an implicit definition requires to introduce axioms and thus
jeopardizes the possibility to extract witnesses from constructive
proofs. However, integrating these axioms in the congruence preserving
normalization, non confusion and the form of the proofs of $N(t)$,
restores this possibility to extract witnesses from constructive
proofs.

\section*{Acknowledgements}

The author wants to thank Ying Jiang for her remarks on a previous draft of 
this note.

\end{document}